\documentstyle [aas2pp4,epsf]{article}

\def\bbox#1{
\relax
\mathchoice
{{\hbox{\boldmath$\displaystyle#1$}}}%
{{\hbox{\boldmath$\textstyle#1$}}}%
{{\hbox{\boldmath$\scriptstyle#1$}}}%
{{\hbox{\boldmath$\scriptscriptstyle#1$}}}%
}

\begin {document}
\title {Filaments and Pancakes in the IRAS $1.2 Jy$ Redshift Catalogue}
\author {B.S. Sathyaprakash$^1$, Varun Sahni$^2$, Sergei Shandarin$^3$, Karl B.
Fisher$^4$\\
$^1$Department of Physics and Astronomy, Cardiff University of Wales,
Cardiff, CF2 3YB, U.K.\\
$^2$Inter-University Centre for Astronomy \& Astrophysics,
Post Bag 4, Ganeshkhind, Pune 411007, India\\
$^3$Department of Physics and Astronomy, University of Kansas,
Lawrence, KS 66045\\
$^4$ Institute for Advanced Studies, Olden lane, Natural Sciences, Bldg
E, Princeton, NJ 08540}

\begin {abstract}

We explore shapes of clusters and superclusters in 
the IRAS $1.2 Jy$ redshift survey with three
reconstructions spanning the range $\beta = 0.1, 0.5, 1.0,$ where $\beta = \Omega^{0.6}/b$, $b$ is the bias factor and $\Omega$ the present value of the dimensionless matter
density. Comparing our results to Gaussian randomized 
reconstructions of the IRAS catalogue, we 
find structures having both planar and filamentary properties. For $\beta = 0.5, ~1.0$
the {\em largest structures in the survey have a distinct tendency to be 
filament-like} in general agreement with the results of N-body simulations.

\end {abstract}

Redshift surveys of galaxies show that the large scale structure of the Universe
has a non-random pattern which on different occasions has been described as 
being cellular, network-like, filamentary, a cosmic web etc. 
(\cite{zesh82,delgh91,bkf96}).
It is also well known that individual structures forming through gravitational 
instability are likely to be anisotropic, since gravitational collapse 
leads generically to collapse along one dimension resulting in the formation of 
pancakes (\cite{sz89,shetal95}), moreover, with the passage of time,  
filamentary 
features acquire greater prominence and the large scale distribution
develops a network like structure
(\cite{ys96,sss96}).

Geometrical properties of large scale structure have evoked great interest in
recent years and mathematical tools as diverse as minimal spanning trees,
genus curves, percolation theory, Minkowski functionals and shape statistics 
have all
been employed in their study. The necessity of using different statistical measures
to study large scale structure arises because traditional indicators
of clustering such as the two-point correlation function, though
robust, neither address the issue of `connectedness' nor 
shape, issues which are central
to an integral understanding of the morphology of large scale structure
and of the physics of gravitational clustering (\cite{sc95}).

Some key issues of the large scale clustering of matter,
such as whether the Great Wall in the North and the Sculptor Wall in the
South are truly one-dimensional `filaments' or are part of a more complex 
cellular structure consisting of sheets and bubbles of which they represent
a limited slice, will be addressed by upcoming
large redshift surveys, such as the Sloan Digital Sky Survey (SDSS) and 
the 2 Degree Field (2dF). In the present paper, we report some 
progress towards this
goal by analysing shapes of clusters and superclusters in the IRAS $1.2 Jy$ 
redshift survey.

Percolation theory, when applied to gravitationally clustered systems, suggests
that most of the matter is likely to be concentrated in 
pancakes, filaments and ribbons, since such an arrangement percolates 
easily ({\it i.e.,} at small values of the filling factor) as borne out 
by N-body simulations which show that the filling factor at
percolation for CDM-type models can be as low as $3$--$5\%$, down from 
$16\%$ for a Gaussian random field (\cite{ks93}). 
Percolation analysis has also shown that the number of distinct clusters in a 
continuous density distribution peaks just before the onset of percolation.
Thus, the percolation transition presents a natural threshold at which to
study the shapes of individual clusters (\cite{sss96,sss98b})
and in our study we shall employ this threshold to study clusters 
in the IRAS survey. (A comprehensive analysis of the IRAS $1.2 Jy$ 
redshift survey using percolation theory was carried out in 
Yess, Shandarin \& Fisher (1997).)

We shall analyse clusters in the IRAS $1.2 Jy$ redshift survey
using a moment-based shape statistic originally suggested for a 
distribution of points by \cite {bs92} (henceforth BS) and modified 
for use on continuous density distributions by \cite{sss96}. We 
briefly describe the modified version of the BS statistic below 
before applying it to clusters obtained at the percolation 
threshold using a nearest neighbors algorithm.

Let $\rho(\bbox x)$ be the density field of matter distribution
defined on a grid with coordinates $\bbox {x}^p = (x^p_1, x^p_2, x^p_3),$ 
$p=1,\ldots,N.$ The first- and second-moments of the density distribution 
around a fiducial point $\bbox {x}^0$ are given by
\begin {equation}
M_i(\bbox x^0; R) = \frac {1}{\cal M} \sum_{p=1}^N
y^p_i \rho\left (\bbox x^p \right )
W\left (|\bbox y^p| \right ),
\end {equation}
\begin {equation}
M_{ij}(\bbox x^0; R) = \frac {1}{\cal M} \sum_{p=1}^N
y^p_i y^p_j \rho\left (\bbox x^p \right )
W\left (|\bbox y^p| \right ),
\end {equation}
where $i,j=1,2,3,$ $R$ is the radius of a window $W$ centred on 
the point $\bbox {x}^0,$ $\bbox y^p\equiv \bbox x^p-\bbox x^0$ 
is the coordinate of the $p$th grid point relative to the fiducial 
point $\bbox x^0$, and $\cal M$ is the total mass in a given region:
\begin {equation}
{\cal M} = \sum_{p=1}^N \rho\left (\bbox x^p \right ) 
W\left (|\bbox x^p-\bbox x^0| \right ).
\end {equation}
In this study we use a spherical top-hat window function.
The moment of inertia tensor $I_{ij}$ can be computed from the moments 
\begin {equation}
I_{ij} = M_{ij} - M_i M_j.
\label{eq:inertia}
\end {equation}
The three eigenvalues, $I_1,$ $I_2$ and $I_3,$ of the inertia
tensor are directly related to
the three principal axes of an ellipsoid fitted to the distribution
of matter around the given point
(fitting by an ellipsoid becomes increasingly accurate in the limit of small
$R$ 
when the distribution can be approximated by a quadratic).
Let us consider the ratios of the eigenvalues arranged in order of increasing 
magnitude, i.e., $I_1>I_2>I_3$: 
\begin {equation}
\mu \equiv \left ({I_2}/{I_1} \right )^{1/2}, \ \ 
\nu \equiv \left ({I_3}/{I_1} \right )^{1/2},
\end {equation}
The BS shape statistic consists of a triad of numbers 
$(S_1, S_2, S_3)$ which can 
be constructed out of the parameters $\mu$ and $\nu$ as follows
\begin {equation}
S_1 = \sin \left [ \frac {\pi}{2} \left (1-\mu \right)^p \right ],\ 
S_2 = \sin \left [ \frac {\pi}{2} a \right ],\ 
S_3 = \sin \left [ \frac {\pi}{2} \nu \right ],
\label {eq:bs1}
\end {equation}
where $p=\log 3/\log 1.5,$ the function $a(\mu,\nu)$ is implicitly
given by
\begin {equation}
\frac {\mu^2}{a^2} - \frac {\nu^2}{a^2(1-\alpha a^{1/3} + \beta a^{2/3})}
= 1,
\label {eq:bs2}
\end {equation}
and $\alpha =  1.9$ and $\beta = - (7/8) 9^{1/3} + \alpha 3^{1/3}.$

\begin {figure}[ht]
\centerline {\epsfxsize 3 true in \epsfbox {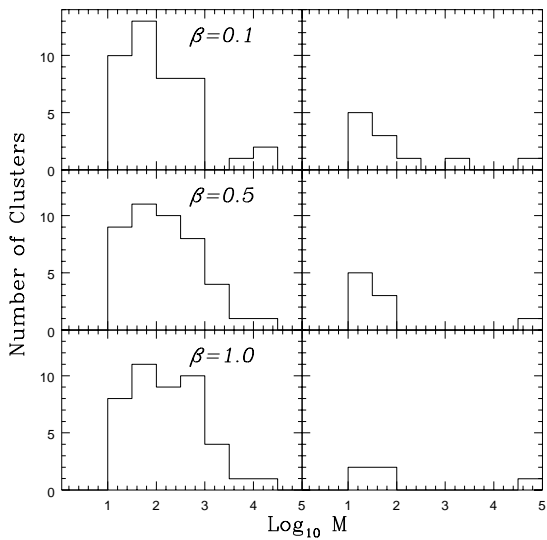}}
\caption{Multiplicity function of clusters in the IRAS 1.2 Jy red-shift 
catalogue (left). Also shown is the multiplicity function for a Gaussian 
randomisation of the IRAS catalogue (right).}
\label{fig:multiplicity.iras}
\end{figure}

\begin {figure}[ht]
\centerline {\epsfxsize 3 true in \epsfbox {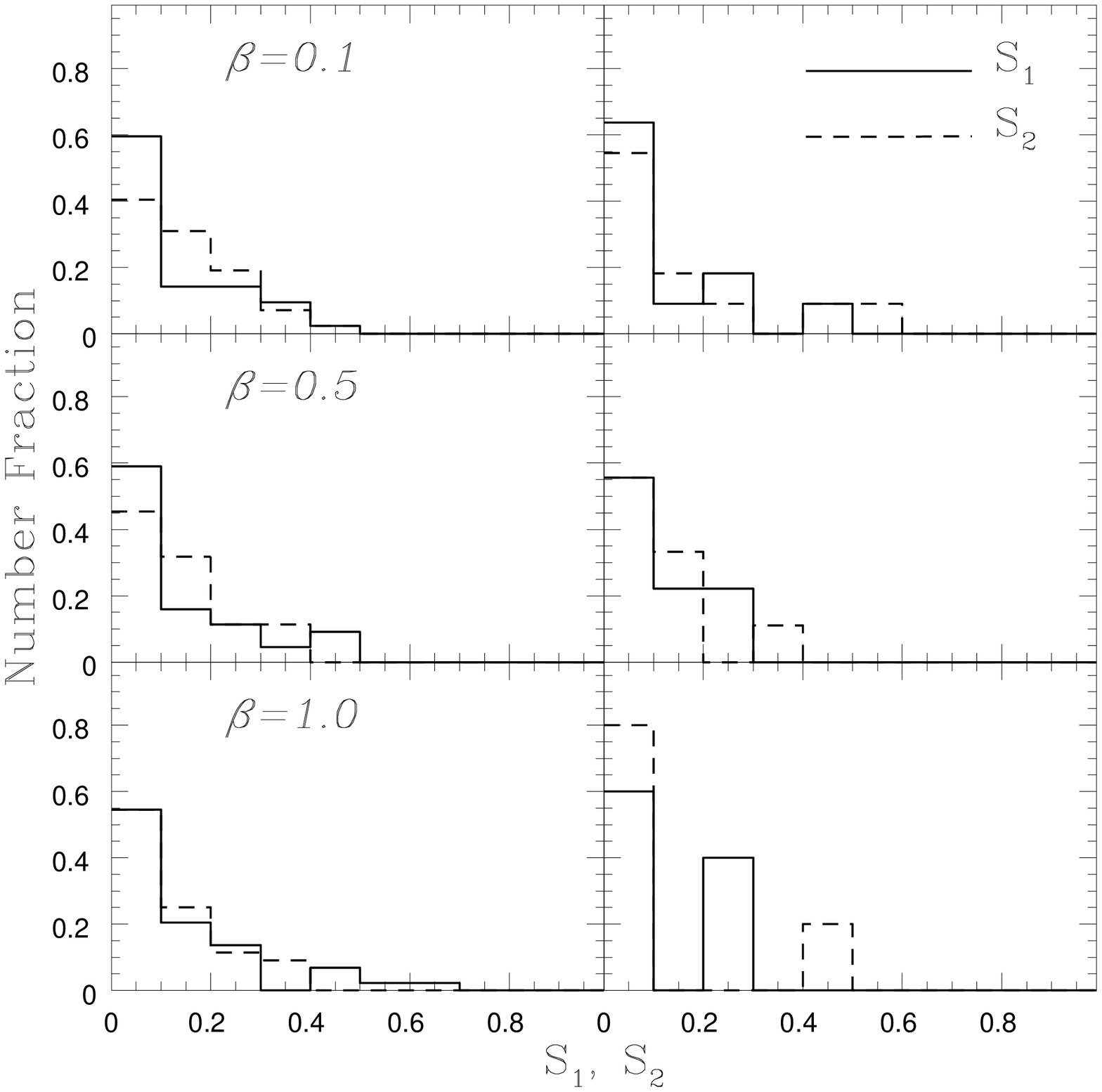}}
\caption{Shape-spectrum of IRAS clusters (left panels) and of the randomized 
catalogue (right panels). }
\label{fig:spectrum.bs.iras}
\end{figure}
As a result $0\le S_i \le 1$, $i = 1,2,3$, {\it i.e.,} the BS statistic can be
thought of as a vector whose components are {\it always} positive and whose
magnitude never exceeds unity. A perfectly spherical distribution has
$I_1=I_2=I_3,$ implying $\mu=\nu=1$, $a=0$ and $(S_1, S_2, S_3)= (0,0,1).$
Similarly, for a planar distribution $I_1=I_2,$ $I_3=0,$ which implies
$\mu=1,$ $\nu=0$ and $a=1,$ so that $(S_1,S_2,S_3) = (0,1,0).$
A distribution which is filamentary has $I_2=I_3=0,$ so that
$\mu=\nu=a=0$ and $(S_1,S_2,S_3)=(1,0,0).$
Thus, the three components of the {\it shape vector} $\bbox {S}= (S_1, S_2, S_3)$ represent filamentarity $S_1$, planarity $S_2$ and sphericity $S_3$.
The magnitude and orientation
of the shape vector describes some morphological properties of a 
distribution, for realistic distributions $S_i\ne 0$, $i = 1,2,3$.
Since the three components of $\bbox S$ are related through 
are depend only on two parameters $\mu$ and $\nu,$ one can consider 
any two of them as being independent. In this work we shall mainly 
work with $S_1$ (linearity) and $S_2$ (planarity), small values of 
these two parameters imply a large value for sphericity $S_3$.

\begin {figure}[ht]
\centerline {\epsfxsize 3 true in \epsfbox {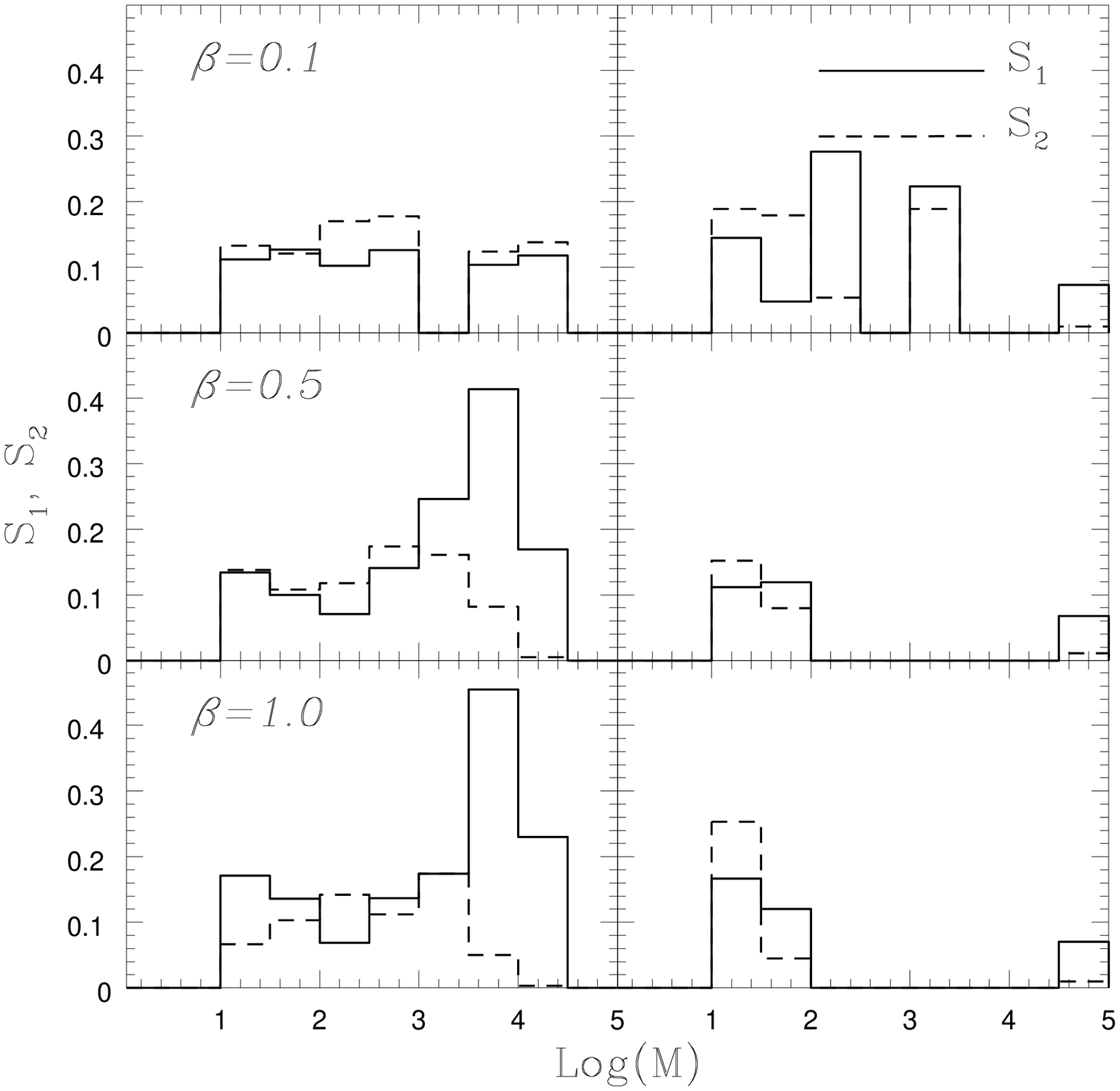}}
\caption {Shape-mass histogram of IRAS clusters (left panels) 
and of the randomized catalogue (right panels).}
\label{fig:shapes.bs.iras}
\end{figure}

We should point out that the moments in
their present form can be used to determine shapes of continuous fields
such as brightness/temperature or density distributions while those
originally defined by BS could not have been used in such cases.
We shall now apply them to study shapes in the IRAS $1.2 ~Jy$ redshift survey.

We now apply the BS shape statistic to overdense 
regions in a Wiener reconstruction of the IRAS $1.2 Jy$ redshift
survey.
The Weiner reconstruction method is useful in cosmology to construct
a real space density field from a galaxy distribution in redshift space
which may be incomplete and sparsely sampled. The Weiner reconstruction
technique uses linear perturbation theory to model and compensate for 
redshift space distortions caused by peculiar velocities of galaxies.
The latter  depend upon the growth rate of the linear density contrast
parametrised by $\beta = \Omega^{0.6}/b$ where $\Omega$ is the present 
value of the cosmic density parameter and $b$ is the linear bias
parameter (for details of Weiner reconstruction see Rybicki \& Press
(1992), Fisher et al. (1995a,b)).
In this study we investigate a set of three reconstructions
of the IRAS density field corresponding to $\beta = 0.1, ~0.5, ~1.0$. 
The real space density field is reconstructed on a $64^3$ grid with side
$200$ $h^{-1}$ Mpc (20,000 km~s$^{-1}$).

Clusters in the IRAS survey are 
identified using percolation theory and a `friends-of-friends'
algorithm on a grid using six nearest neighbors.
A generic feature of any continuous density distribution (and therefore true also of
the reconstructed IRAS density field) is that at very
high thresholds only a small volume is in the overdense phase, so that the
resulting number of clusters is very small and so is the volume in the largest cluster.
As the density threshold is lowered, the number of clusters increases and the
volume in the largest cluster grows rapidly due to merging of nearby clusters.
At a critical value of the density (the {\it percolation threshold}) 
the largest cluster
percolates, spanning the entire region of interest in a homogeneous sample. 
Further lowering of the threshold increases mergers, resulting in
a subsequent decrease in the
number of clusters.
The total number of clusters, thus, peaks at thresholds close to
percolation, making the percolation 
transition an objective and useful density threshold at which to study
properties of individual clusters (\cite{sss96,sss98b}),
and we shall use this threshold for studying cluster 
shapes in the IRAS survey.

The cluster multiplicity function for three reconstructions of the IRAS density
field $\beta = 0.1, 0.5, 1.0$, is
shown in Fig.~\ref{fig:multiplicity.iras} (left panels). 
Also shown are Gaussian randomized reconstructions of the IRAS catalogue, 
which serve as useful standards for our analysis (Fig.~\ref{fig:multiplicity.iras} 
right panels).
In the IRAS catalogue there are between 40--50 
clusters in each reconstruction. About half of them are small clusters
a third are of intermediate mass and the rest are very massive.
In a low $\Omega$-Universe, or if the bias is very large, the clusters
tend to be not as massive as in a high $\Omega$-Universe or if there
is no biasing. The randomized IRAS catalogue contains many small clusters, 
in Fig.~\ref{fig:multiplicity.iras} we only show clusters having volume 
larger than 8 grid 
cells, consequently, the randomized catalogue is seen to contain
fewer large clusters than 
the IRAS catalogue, even though the total number of clusters in both catalogues is 
roughly the same. (Details of Gaussian randomisation of the IRAS density field 
can be found in \cite{ysf97}.) 

In Fig.~\ref{fig:spectrum.bs.iras} we show the `shape-spectrum' -- 
the number fraction of IRAS clusters   
having a given value of the Babul and Starkman shape 
parameters $S_1$ \& $S_2$. From the shape-spectrum we 
see that most IRAS clusters are predominantly 
isotropic/spherical, however, a small fraction can have fairly 
large amounts of planarity/filamentarity. 
Fig.~\ref{fig:spectrum.bs.iras} does not provide a comprehensive picture since 
it weighs all clusters equally regardless of whether they are large or small.
To explore the mass dependence of shape parameters we show the
shape-mass histogram of IRAS density fields 
in Fig. \ref{fig:shapes.bs.iras}. We find that for $\beta = 0.1$ 
clusters in all mass ranges have almost equal amounts of planarity and filamentarity.
However, for $\beta = 0.5, ~ 1.0$, the largest clusters (superclusters)
tend to be predominantly filamentary, in agreement with the 
results of N-body simulations (\cite{sss96,sss98b}). (The 
statistical significance of these results is not yet clear due to the
small number of such (large) objects in the IRAS survey volume.) 
Table~\ref{table:iras} is our
catalogue of clusters in the IRAS
density fields reconstructed with $\beta=0.5.$ 
(We do not include very small clusters having volume less than 8 grid cells.) 
We find several large clusters with significant amounts of filamentarity/planarity. 
It should be pointed out that the largest `percolating' supercluster is likely to be 
`tree-like' with several branches emanating from a `central trunk'. This will
give it an isotropic appearance on very large scales. Since the BS statistic is 
moment-based it is likely to interpret such an isotropic structure as being
spherical ! This short coming of moment-based shape statistics can be avoided
if one works with shape diagnostics based on Minkowski functionals as 
demonstrated in \cite{sss98a} and \cite{sss98b}.
(In addition, clusters occuring at the edge of the box may show enhanced
planarity because of boundary effects.)

To conclude,
we have addressed the issue of morphology of clusters and superclusters
in the IRAS $1.2~ Jy $ redshift catalogue. We find that individual 
clusters defined at the percolation threshold can have significant 
amounts of both filamentarity and 
planarity, the largest clusters appearing to be strongly filamentary.
Although these results are broadly in agreement with
recent studies of N-body simulations, their statistical implications are 
still unclear, mainly because of the sparseness of the IRAS catalogue. 
However, the stage is now set
to analyse larger and deeper three-dimensional redshift surveys 
complementing the IRAS survey, such as the 2 Degree Field (2dF) and
the Sloan Digital Sky Survey (SDSS).
A comprehensive study of cluster and supercluster shapes in these surveys
is bound to shed more light on the abundance of pancakes, filaments and ribbons
and on the geometry of large scale structure, whether bubble-like,
network-like or some other !

{\small
\baselineskip 12 true pt

}

\begin {table}
\begin {center}
\caption{\small
Catalogue of clusters/superclusters from the IRAS reconstruction with
$\beta = 0.5$. The list contains many interesting
structures, some of them quite massive 
with significant amounts of planarity and/or filamentarity.
(The first column is the cluster number ($n$), the second is volume in grid
units ($V$, and the third mass/$10^2$ ($M$)}
\bigskip
\small
\begin {tabular} {rrrc}
\hline
 $n$ & $V$ & $M$ & $(S_1,     S_2,     S_3)$ \\
\hline
1&10927 & 238~~~      & (0.169,   0.005,   0.684)\\
2&~1898 &  36~~~      & (0.413,   0.082,   0.375)\\
3&~1417 &  27~~~      & (0.007,   0.388,   0.503)\\
4&~~905 &  17~~~      & (0.493,   0.024,   0.394)\\
5&~~775 &  15~~~      & (0.064,   0.221,   0.517)\\
6&~~494 &  10~~~      & (0.419,   0.012,   0.465)\\
7&~~302 &   5.3~      & (0.266,   0.199,   0.372)\\
8&~~302 &   5.1~      & (0.204,   0.266,   0.367)\\
9&~~287 &   4.9~      & (0.015,   0.104,   0.712)\\
10&~~232 &   4.4~      & (0.113,   0.128,   0.541)\\
11&  252 &   4.2~      & (0.244,   0.271,   0.339)\\
12&~~239 &   4.2~      & (0.200,   0.024,   0.597)\\
13&~~208 &   3.5~      & (0.022,   0.376,   0.471)\\
14&~~197 &   3.3~      & (0.067,   0.027,   0.728)\\
15&~~161 &   3.0~      & (0.010,   0.051,   0.799)\\
16&~~165 &   2.8~      & (0.064,   0.007,   0.795)\\
17&~~123 &   2.0~      & (0.003,   0.107,   0.760)\\
18&~~117 &   1.9~      & (0.066,   0.189,   0.539)\\
19&~~108 &   1.8~      & (0.011,   0.293,   0.557)\\
20&~~~97 &   1.6~      & (0.047,   0.114,   0.635)\\
21&~~~97 &   1.6~      & (0.012,   0.017,   0.862)\\
22&~~~93 &   1.5~      & (0.280,   0.020,   0.540)\\
23&~~~85 &   1.4~      & (0.090,   0.239,   0.473)\\
24&~~~73 &   1.2~      & (0.122,   0.148,   0.515)\\
25&~~~48 &   0.78      & (0.070,   0.188,   0.535)\\
26&~~~38 &   0.78      & (0.006,   0.016,   0.886)\\
27&~~~44 &   0.74      & (0.061,   0.019,   0.757)\\
 28&~~~37 &   0.59      & (0.208,   0.180,   0.422)\\
 29&~~~31 &   0.56      & (0.120,   0.016,   0.691)\\
 30&~~~32 &   0.53      & (0.399,   0.011,   0.482)\\
 31&~~~25 &   0.40      & (0.002,   0.180,   0.698)\\
 32&~~~24 &   0.39      & (0.047,   0.179,   0.573)\\
 33&~~~21 &   0.38      & (0.031,   0.047,   0.748)\\
 34&~~~23 &   0.37      & (0.035,   0.349,   0.465)\\
 35&~~~21 &   0.35      & (0.124,   0.003,   0.745)\\
 36&~~~17 &   0.27      & (0.052,   0.379,   0.422)\\
 37&~~~17 &   0.27      & (0.060,   0.024,   0.746)\\
 38&~~~16 &   0.27      & (0.320,   0.004,   0.569)\\
 39&~~~16 &   0.27      & (0.008,   0.040,   0.826)\\
 40&~~~15 &   0.25      & (0.482,   0.066,   0.351)\\
 41&~~~14 &   0.22      & (0.111,   0.190,   0.491)\\
 42&~~~11 &   0.18      & (0.000,   0.117,   0.799)\\
 43&~~~10 &   0.16      & (0.087,   0.117,   0.579)\\
 44&~~~ 9 &   0.14      & (0.083,   0.302,   0.437)\\
\hline
\end {tabular}
\end {center}
\label{table:iras}
\end {table}
\end{document}